\documentstyle[12pt]{article}
\def\ii{\'{\char'20}}

\begin{document}

\newcommand{\ep}{\epsilon}
\newcommand{\fr}{\frac}
\newcommand{\reals}{\mbox{${\rm I\!R }$}}
\newcommand{\nats}{\mbox{${\rm I\!N }$}}
\newcommand{\intgs}{\mbox{${\rm Z\!\!Z }$}}
\newcommand{\cam}{{\cal M}}
\newcommand{\caz}{{\cal Z}}
\newcommand{\cao}{{\cal O}}
\newcommand{\cac}{{\cal C}}
\newcommand{\aaa}{\int\limits_{mR}^{\infty}dk\,\,}
\newcommand{\bbb}{\left[\left(\frac k R\right)^2-m^2\right]^{-s}}
\newcommand{\ccc}{\frac{\partial}{\partial k}}
\newcommand{\fff}{\frac{\partial}{\partial z}}
\newcommand{\iikma}{\aaa \bbb \ccc}
\newcommand{\ddd}{\int\limits_{mR/\nu}^{\infty}dz\,\,}
\newcommand{\eee}{\left[\left(\frac{z\nu} R\right)^2-m^2\right]^{-s}}
\newcommand{\lll}{\frac{(-1)^j}{j!}}
\newcommand{\iinma}{\ddd\eee\fff}
\newcommand{\cah}{{\cal H}}
\newcommand{\nn}{\nonumber}
\renewcommand{\theequation}{\mbox{\arabic{section}.\arabic{equation}}}
\newcommand{\komplex}{\mbox{${\rm I\!\!\!C }$}}
\newcommand{\sip}{\frac{\sin (\pi s)}{\pi}}
\newcommand{\numr}{\left(\frac{\nu}{mR}\right)^2}
\newcommand{\mzs}{m^{-2s}}
\newcommand{\rzs}{R^{2s}}
\newcommand{\abl}{\partial}
\newcommand{\g}{\Gamma\left(}
\newcommand{\zzz}{\int\limits_{\gamma}\frac{dk}{2\pi i}\,\,}
\newcommand{\yyy}{(k^2+m^2)^{-s}\frac{\partial}{\partial k}}
\newcommand{\ikma}{\zzz \yyy}
\newcommand{\ead}{e_{\alpha}(D)}
\newcommand{\sual}{\sum_{\alpha =1}^{D-2}}
\newcommand{\sulnu}{\sum_{l=0}^{\infty}}
\newcommand{\sujnu}{\sum_{j=0}^{\infty}}
\newcommand{\suani}{\sum_{a=0}^i}
\newcommand{\suanzi}{\sum_{a=0}^{2i}}
\newcommand{\zend}{\zeta_D^{\nu}}
\newcommand{\amed}{A_{-1}^{\nu ,D}(s)}
\renewcommand{\and}{A_{0}^{\nu ,D}(s)}
\newcommand{\aid}{A_{i}^{\nu ,D}(s)}
\def\beq{\begin{eqnarray}}
\def\eeq{\end{eqnarray}}

\begin{titlepage}

\title{\begin{flushright}
{\normalsize UB-ECM-PF 95/3 }
\end{flushright}
\vspace{3mm}
{\Large \bf Heat-kernel coefficients of the Laplace operator on the
$D$-dimensional ball}}
\author{M. Bordag\thanks{E-mail address:
bordag@qft.physik.uni-leipzig.d400.de}\\
Universit{\"a}t Leipzig, Institut f{\"u}r Theoretische Physik,\\
Augustusplatz 10, 04109 Leipzig, Germany\\
E. Elizalde\thanks{E-mail address: eli@zeta.ecm.ub.es}\\
CEAB, CSIC, Cam\'{\i} de Santa
B\`arbara,
17300 Blanes,
\\ and Departament d'ECM and IFAE, Facultat de F{\ii}sica,\\
Universitat de Barcelona, Av. Diagonal 647, 08028 Barcelona\\
 K. Kirsten\thanks{Alexander von Humboldt foundation
fellow, E-mail address: klaus@zeta.ecm.ub.es}\\
Departament d'ECM, Facultat de F{\ii}sica,
\\ Universitat de Barcelona, Av. Diagonal 647, 08028 Barcelona \\
Spain}

\thispagestyle{empty}

\vspace*{-1mm}

\maketitle

\vspace*{-2mm}

\begin{abstract}
We present a very quick and powerful method for the calculation
of heat-kernel coefficients. It makes use of rather common ideas, as
integral representations of the spectral sum, Mellin transforms,
non-trivial commutation of series and integrals and skilful
analytic continuation of zeta functions on the complex plane.
We
apply our method to the case of the heat-kernel expansion of the Laplace
operator
on a $D$-dimensional ball with either Dirichlet, Neumann or, in general,
Robin boundary conditions. The final formulas are quite simple. Using
this case as an example, we illustrate in detail our
 scheme ---which serves for the calculation
of an (in principle) arbitrary number of heat-kernel coefficients in
any situation when the basis functions are known. We provide a
complete list of new results
for the coefficients $B_3,...,B_{10}$, corresponding to the
$D$-dimensional ball with all the mentioned boundary conditions and
$D=3,4,5$.
\end{abstract}
\end{titlepage}
\section{Introduction}
\setcounter{equation}{0}
An important issue for more than twenty years now has been
 to obtain explicitly the coefficients which
appear in the
short-time expansion of the heat-kernel $K(t)$ corresponding to a
 Laplacian-like
operator on a $D$-dimensional manifold $\cam$. In mathematics this
interest stems, in particular, from the well-known connections that
exist between the heat-equation
and the Atiyah-Singer index theorem \cite{gilkey84}. In physics,
the importance of that expansion is notorious in different
domains of quantum
field theory,  where it is commonly known as the (integrated)
Schwinger-De Witt proper-time expansion \cite{birelldavies82,dewitt75}.
In this context, the heat-equation for an elliptic (in general
pseudoelliptic) differential operator
$P$ and the corresponding zeta function $\zeta_P (s)$ has been realized
to be a particularly useful tool for the determination of effective
actions \cite{hawking77} and for the calculation of vacuum or Casimir
energies \cite{blauvisserwipf88} (a fundamental issue for understanding
the vacuum structure of a quantum field theory). Here usually
the derivative
$\zeta'_P(0)$ of the zeta function \cite{hawking77} and its
value at $s=-1/2$ (sometimes the principal part) are
needed \cite{blauvisserwipf88,dolannash92}.

In this paper we would like to exploit another property of the zeta
function
$\zeta_P (s)$ corresponding to an elliptic operator $P$, namely its
well-known close connection with the heat-kernel expansion. In spite of
the fact that almost everybody is aware of such connection, its actual
use in the literature has remained very scarce till now.
If the manifold $\cam$ has a boundary $\partial \cam$, the coefficients
$B_n$ in the short-time expansion have both a volume and a boundary
part \cite{greiner71,gilkey76}. It is usual to write this
expansion in the form
\beq
K(t) \sim (4\pi t)^{-\frac D 2}\sum_{k=0,1/2,1,...}^{\infty}
B_kt^k, \label{eq:1.1}
\eeq
with
\beq
B_k =\int_{\cam}dV\,\,b_n+\int_{\partial \cam}dS\,\,c_n.\label{eq:1.2}
\eeq
For the volume part very effective systematic schemes have been
developed (see for example
\cite{avramidi91,fullingkennedy88,amsterdamskiberkinoconnor89}).
The calculation of $c_n$, however, is in general more difficult.
Only quite recently has the coefficient $c_2$ for Dirichlet and
for Neumann boundary conditions been found [12-17].
\nocite{bransongilkey90,mossdowker89,mcavityosborn91,dettkiwipf92}
\nocite{dowkerschofield90,cognolavanzozerbini90}
Very new results on the coefficient $B_{5/2}$ for manifolds with totally
geodesic boundaries will be given in \cite{gilkeynew}.

When using the general formalism
of Ref.~\cite{bransongilkey90} for higher-spin particles, Moss and
Poletti \cite{mosspoletti90,mosspoletti90a} found a discrepancy
with the direct calculations of D'Eath and Esposito
\cite{deathesposito91}
(see also [22-25]).
\nocite{esposito94,espositokamenshchikmishakovpollifrone94}
\nocite{espositokamenshchikmishakovpollifrone94a,esposito94a}
The latter results have been confirmed in
\cite{kamenshchikmishakov92,barvinskykamenshchikkarmazin92},
where a new systematic scheme for the calculation of $c_2$ has
been designed in the context of the Hartle-Hawking wave-function of
the universe and for the case when the whole set of basis functions
is known \cite{kamenshchikmishakov92,barvinskykamenshchikkarmazin92}.
Finally, very recently the discrepancy has been resolved completely
\cite{vassilevich95} and now the results that are found using the general
algorithm \cite{mosspoletti94} are in agreement with
those coming from the direct calculations [21-27].
\nocite{deathesposito91,esposito94,espositokamenshchikmishakovpollifrone94}
\nocite{espositokamenshchikmishakovpollifrone94a}
\nocite{esposito94a,kamenshchikmishakov92,barvinskykamenshchikkarmazin92}

The connection between the heat-kernel expansion, Eq.~(\ref{eq:1.1}) and
the associated zeta function is established through the formulas
\cite{voros87}
\beq
Res\,\,\zeta (s) = \frac{B_{\frac m 2 -s}}{(4\pi )^{\frac m 2} \Gamma
(s)},\label{residuum}
\eeq
for $s=\frac m 2,\frac{m-1}2,...,\frac 1 2;-\frac{2l+1}2,$ for $l\in
\nats_0$, and
\beq
\zeta (-p) = (-1)^p p! \frac{B_{\frac m 2 +p}}{(4\pi )^{\frac m 2}},
\label{fusvalue}
\eeq
for $p\in \nats _0$.
The aim of the present article is to show that these equations,
(\ref{residuum}) and (\ref{fusvalue}),
can actually serve as a very convenient starting point for the calculation of
th
coefficients $B_k$, even in the cases when the eigenvalues of the
operator $P$ under consideration are not known. The good knowledge in
explicit zeta-function evaluations that have been accumulated in the
past few years (for a review of many results in this respect, see
\cite{eorbz}) will allow us to elaborate a very competitive method of
calculation of the heat-kernel coefficients
which makes use of rather common ingredients, such as
integral representations of the spectral sum, Mellin transforms,
non-trivial commutation of series and integrals and skilful
analytic continuation of zeta functions on the complex plane.

To explain the method in detail we
will consider the Laplace operator on the $D$-dimensional ball with
Dirichlet, Neumann or (in general) Robin boundary conditions. Earlier
investigations
on the first few coefficients are due, for $D=1$, to Stewartson and
Waechter
\cite{stewartsonwaechter71}, to Waechter in $D=2$ \cite{waechter72} and
to Kennedy \cite{kennedy78,kennedy78a} in up to $D=5$ dimensions (for
recent results on the functional determinant of the Laplace operator on
the three and four dimensional ball see \cite{dowker}).
In these references the method was based on the use of
Laplace transformations of the heat-kernel $K(t)$ itself. In that method
an intermediate cut off has to be introduced at some point
---because one needs to consider the Laplace transform of a
 function which is singular
at $t=0$. In contrast, in our approach it is the complex argument $s$
of the
zeta function of the Laplace operator which very neatly serves for the
regularization of all sums (in just the usual way \cite{eorbz}).

The layout of the paper is as follows. In section 2 we briefly
describe the eigenvalue problem of the massive Laplace operator on the
ball and derive a representation of the associated zeta function in
terms of a contour integral. We consider the {\it massive} Laplace
operator because the analytical continuation procedure is slightly
easier for the case of non-vanishing mass. In section 3 we describe how
an analytical representation of the zeta function ---valid in the strip
$(1-N)/2<\Re s <1$--- can be obtained for any $N$, restricting our
considerations
in this section to $D=3$ and to the case of Dirichlet boundary
conditions.
This representation will display very clearly the meromorphic structure
of
the zeta function. As is then shown in section 4,  from this
representation it is quite immediate to read off
special properties, as the ones reflected by
 (\ref{residuum}) and (\ref{fusvalue}), in
order to find the heat-kernel coefficients. In section 5 we explain the
small changes in the procedure that are necessary in order to treat
Robin boundary conditions, in general. Finally,
in section 6 we study the modification to be introduced in the
formulas for considering any arbitrary dimension $D$. In appendix A we
exhibit some technical details of the calculation and in
Apps.~B, C and D we give explicit tables of the heat-kernel
coefficients for Dirichlet, Neumann
and general Robin boundary conditions, for the dimensions
$D=3,4,5$.

\section{Heat-kernel coefficients on the $D$-dimensional ball}
\setcounter{equation}{0}
As explained in the introduction, we are interested in the zeta function
of the operator $(-\Delta +m^2)$ on the $D$-dimensional ball
$B^D=\{x\in\reals ^D; |x|\leq R\}$ endowed with Dirichlet, Neumann or
Robin boundary conditions. The zeta function is formally defined as
\beq
\zeta (s) =\sum_k \lambda_k ^{-s},\label{eq:2.1}
\eeq
with the eigenvalues $\lambda_k$ being determined through
\beq
(-\Delta +m^2) \phi_k (x) =\lambda_k \phi _k (x)\label{eq:2.2}
\eeq
($k$ is in general a multiindex here), together with one of the
 three boundary
conditions above.
It is convenient to introduce a spherical coordinate basis, with $r=|x|$
and $D-1$
angles $\Omega =(\theta_1 ,...,\theta_{D-2},\varphi )$.
In these coordinates, a complete set of solutions of Eq.~(\ref{eq:2.2})
together with one of the mentioned boundary conditions may be given in
the form
\beq
\phi_{l,m,n}(r,\Omega )=r^{1-\frac D 2} J_{l+\frac{D-2}2} (w_{l,n}r)
Y_{l+\frac D 2} (\Omega ),
\eeq
with $J_{l+(D-2)/2}$ being Bessel functions and
$Y_{l+D/2}$ hyperspherical harmonics
\cite{erdelyimagnusoberhettingertricomi53}. The $w_{l,n}$
$(>0)$ are determined through the boundary conditions by
 \beq
J_{l+\frac{D-2} 2 }(w_{l,n}R) &=& 0, \phantom{a}{\rm for}\phantom{a}
{\rm Dirichlet}\phantom{a}{\rm boundary}\phantom{a}{\rm
conditions},\nn\\
\frac u R J_{l+\frac{D-2} 2 }(w_{l,n}R) +
w_{l,n} J'_{l+\frac{D-2} 2 }(w_{l,n}r)\left|_{r=R} \right.&=&
0, \phantom{a}{\rm for}\phantom{a}{\rm Robin}\phantom{a}{\rm
boundary}\phantom{a}{\rm conditions.}\label{eq:2.3a} \eeq
As is clear, the case
$u=(1-D/2)$ of the (general) Robin boundary conditions
corresponds to the Neumann boundary conditions.
In this notations, using $\lambda_{l,n}=w_{l,n}^2+m^2$, the zeta
function can be given in the form \beq
\zeta (s) =\sum_{n=0}^{\infty} \sulnu d_l (D) (w_{l,n}^2
+m^2)^{-s},\label{eq:2.3}
\eeq
where  $w_{l,n} \ (>0)$ is defined as the n-th root of the l-th
equation.
Here the sum over $n$ is extended over all possible roots  $w_{l,n}$
on the positive real axis, and $d_l (D)$ is the number of independent
harmonic polynomials, which defines the degeneracy of each value of $l$
and $n$ in $D$ dimensions. Explicitly,
\beq
d_l (D) = (2l+D-2) \frac{(l+D-3)!}{l!\, (D-2)!}.\label{eq:2.4}
\eeq
Furthermore, here and in what follows the prime will always mean
derivative of the function with respect to its argument.

To distinguish in the notation among the different cases,
 we will use the indices D, N and R to denote Dirichlet,
Neumann
and Robin boundary conditions, respectively. Thus, we will write
$\zeta_D$, $\zeta_N$
and $\zeta_R$ for the corresponding zeta functions. Using for
the moment
the unified notation $\Phi_{l+(D-2)/2}(w_{l,n}R) =0$ for the boundary
condition Eq.~(\ref{eq:2.3a}), it turns out that
Eq.~(\ref{eq:2.3}) may be written under the
form of a contour integral on the complex plane,
\beq
\zeta (s)= \sulnu d_l (D) \ikma\ln
\Phi_{l+\frac{D-2}2}(kR),\label{eq:2.5} \eeq
where the contour $\gamma$ runs counterclockwise and must enclose all
the solutions
of (\ref{eq:2.3a}) on the positive real axis (for a similar treatment of
the zeta function as a contour integral see
\cite{kamenshchikmishakow92,barvinskykamenshchikkarmazin92,bordag95}).
This representation of the zeta function in terms of a contour integral
around some circuit $\gamma$ on the complex plane, Eq.~(\ref{eq:2.5}),
is the {\it first step} of our procedure.

Depending on the value of the
dimension $D$ and on the boundary conditions chosen,
the analysis of the zeta function,
Eq.~(\ref{eq:2.5}) ---to be given below--- will differ, but
just in small details.  For this reason, we
will only describe at length the case of the three-dimensional ball with
Dirichlet
boundary condition. The derivation of the analogous results for the
other boundary conditions and higher dimensions will then be clear, and
shall be indicated only briefly.

\section{A quick procedure for calculating  heat-kernel coefficients}
\setcounter{equation}{0}
As explained above, we will illustrate the procedure in the case of the
three-dimensional ball with Dirichlet boundary conditions. For $D=3$
the degeneracy is $d_l (3) =2l+1$, so that the starting
point of the calculation reads (we omit further indication of the
dimension in the notation)
\beq
\zeta_D(s) =\sulnu (2l+1) \ikma\ln J_{l+\frac 1 2} (kR).\label{eq:3.1}
\eeq
As it stands, the representation (\ref{eq:3.1}) is valid for $\Re s
>3/2$. However, we are interested in the properties of $\zeta_D(s)$ in
the range $\Re s < 0$ and
thus, we need to perform the analytical continuation to the left domain
of the complex plane.
Before considering in detail the $l$-summation, we will first proceed
with the $k$-integral alone.

The first specific idea is to shift the integration contour and place it
along the imaginary
axis. In order to avoid contributions coming from the origin $k=0$, we
will consider (with $\nu =l+1/2$) the expression
\beq
\zend = \ikma \ln \left(k^{-\nu} J_{\nu} (kR) \right),\label{eq:3.2}
\eeq
where the additional factor $k^{-\nu}$ in the logarithm
does not change the result, for no additional pole is enclosed.
One then easily obtains
\beq
\zend =\sip \int\limits_m^{\infty}dk\,\,
[k^2-m^2]^{-s}\frac{\partial}{\partial k}
\ln \left(k^{-\nu }I_{\nu} (kR) \right)\label{eq:3.3} \eeq
valid in the strip $1/2 <\Re s <1$. A similar representation valid for
$m=0$ has been given in
\cite{elizaldeleseduarteromeo93,leseduarteromeo94}.

As the {\it second step} of our method,
we make use of the uniform expansion
of the Bessel function $I_{\nu} (k)$ for $\nu \to \infty$ as $z=k/\nu$
fixed \cite{abramowitzstegun72}. One has
\beq
I_{\nu} (\nu z) \sim \frac 1 {\sqrt{2\pi \nu}}\frac{e^{\nu
\eta}}{(1+z^2)^{\frac 1 4}}\left[1+\sum_{k=1}^{\infty} \frac{u_k (t)}
{\nu ^k}\right],\label{eq:3.4}
\eeq
with $t=1/\sqrt{1+z^2}$ and $\eta =\sqrt{1+z^2}+\ln
[z/(1+\sqrt{1+z^2})]$.
The first few coefficients are listed in \cite{abramowitzstegun72},
higher coefficients are immediate  to obtain by using the recursion
\cite{abramowitzstegun72}
\beq
u_{k+1} (t) =\frac 1 2 t^2 (1-t^2) u'_k (t) +\frac 1 8 \int\limits_0^t
d\tau\,\, (1-5\tau^2 ) u_k (\tau ),\label{eq:3.5}
\eeq
starting with $u_0 (t) =1$. As is clear, all the $u_k (t)$ are
polynomials in $t$. Furthermore, the coefficients $D_n (t)$ defined by
\beq
\ln \left[1+\sum_{k=1}^{\infty} \frac{u_k (t)}{\nu ^k}\right] \sim
\sum_{n=1}^{\infty} \frac{D_n (t)}{\nu ^n}\label{eq:3.6}
\eeq
are easily found with the help of a simple computer program.

Now comes what can be considered as the {\it third step} of our method.
By adding and subtracting $N$ leading terms of the asymptotic
expansion, Eq.~(\ref{eq:3.6}), for $\nu \to \infty$, Eq.~(\ref{eq:3.3})
may be split into the following pieces
\beq
\zend =Z_D^{\nu}(s) +\sum_{i=-1}^{N}A_i^{\nu ,D}(s),\label{eq:3.7}
\eeq
with the definitions
\beq
Z_D^{\nu}(s) &=& \sip \iinma\left\{\ln\left[z^{-\nu}I_{\nu} (z\nu
)\right]\right.\label{eq:3.8}\\
& &\hspace{3cm}\left.
-\ln\left[\frac{z^{-\nu}}{\sqrt{2\pi\nu}}\frac{e^{\nu\eta}}
{(1+z^2)^{\frac
1 4}}\right]-\sum_{n=1}^N \frac{D_n (t)}{\nu ^n}\right\},\nn
\eeq
and
\beq
A_{-1}^{\nu ,D} &=& \sip \iinma \ln \left(z^{-\nu}
e^{\nu\eta}\right),\label{eq:3.9a}\\
A_{0}^{\nu ,D} &=& \sip \iinma \ln
(1+z^2) ^{-\frac 1 4},\label{eq:3.9}\\
A_{i}^{\nu ,D} &=& \sip \iinma
\left(\frac{D_i (t)}{\nu ^i}\right).\label{eq:3.9b}
\eeq
The essential idea is conveyed here by the fact that the representation
(\ref{eq:3.7}) has
the following important
properties. First, by considering the asymptotics of the
integrand in Eq.~(\ref{eq:3.8}) for $z\to mR/\nu$ and $z\to\infty$, it
can be seen that the function
\beq
Z_D(s) =\sulnu (2l+1) Z_D^{l+\frac 1 2} (s)\nn
\eeq
is analytic on the strip $(1-N)/2<\Re s <1$. For this reason, it gives
no
contribution to the residue of $\zeta_D (s)$ in that strip. Furthermore,
for $s=-k$, $k\in\nats_0$, $k<-1+N/2$, we have $Z(s) =0 $ and, thus, it
also yields no contribution to the values of the zeta function at these
points. Together with Eqs.~(\ref{residuum}) and (\ref{fusvalue}), this
result means that the heat-kernel coefficients are just determined
by the terms $A_i^D(s)$ with
\beq
A_i ^D (s) =\sulnu (2l+1) A_i ^{l+\frac 1 2,D} (s).\label{eq:3.10}
\eeq
As they stand, the $\aid$ in Eqs.~(\ref{eq:3.9a}), (\ref{eq:3.9}) and
(\ref{eq:3.9b}) are well
defined on the strip $1/2<\Re s<1$ (at least). And we will now show
that the analytic
continuation in the parameter $s$ to the whole of the complex plane, in
terms
of known functions, can be performed. Keeping in mind that $D_i (t)$ is
a polynomial in $t$, all the $\aid$ are in fact hypergeometric
functions, which is seen by means of the basic relation
\cite{gradshteynryzhik65} \beq
_2F_1 (a,b;c;z) =\frac{\Gamma (c)}{\Gamma (b)\Gamma (c-b)}\int_0^1dt\,\,
t^{b-1} (1-t)^{c-b-1}(1-tz)^{-a}.\nn
\eeq
Let us consider first in detail $\amed$, $\and$, and the
corresponding $A_{-1}^D(s)$, $A_0^D(s)$. One finds immediately that
\beq
\amed &=& \frac{\mzs}{2\sqrt{\pi}}Rm\frac{\g s-\frac 1 2 \right)}{\Gamma
(s)}~_2F_1 \left(-\frac 1 2,s-\frac 1 2;\frac 1
2;-\numr\right)-\frac{\nu}2 \mzs ,\label{eq:3.11}\\
\and &=& -\frac 1 4 \mzs ~_2F_1 \left(1,s;1,-\numr\right) =-\frac 1 4
m^{-2s} \left[1+\numr\right]^{-s},\label{eq:3.12}
\eeq
where in the last equality we have used that $_2F_1 (a,s;a;x)
=(1-x)^{-s}$.

The
next step is to consider the summation over $l$. For $\amed$
this is best done using a Mellin-Barnes type integral representation
of the hypergeometric functions
\beq
_2F_1 (a,b;c;z) =\frac{\Gamma (c)}{\Gamma (a)\Gamma (b)}\frac 1 {2\pi i}
\int\limits_{\cac} dt\,\,\frac{\Gamma (a+t)\Gamma (b+t) \Gamma
(-t)}{\Gamma (c+t)}(-z)^{t},\label{eq:3.13}
\eeq
where the contour is such that the poles of $\Gamma (a+t) \Gamma (b+t)
/\Gamma (c+t)$ lie to the left of it and the poles of $\Gamma (-t)$ to
the right \cite{gradshteynryzhik65}. After
interchanging
the summation over $l$ and the integration in (\ref{eq:3.13}), the
result will be a Hurwitz zeta function, which is defined as
\beq
\zeta_H (s;v) =\sulnu (l+v)^{-s},  \ \ \ \  \Re s > 1.\label{eq:3.14}
\eeq
However, as is well known, one has to be very careful with this kind of
manipulations, what has been realized and explained with great detail in
\cite{weldon86,elizalderomeo89a,elizaldekirstenzerbini95}. This point
 is of crucial importance (it has been the source of many errors in
the literature over the past ten years \cite{eorbz}) and can be
considered as the {\it fourth step} of our original procedure here.
Applying the method, as described in the mentioned references,
 to $A_{-1}^D(s)$,
\beq
A_{-1}^D(s) &=& \sulnu (2l+1) \left[\frac{\mzs}{2\sqrt{\pi}}Rm\frac{\g
s-\frac 1 2\right)}{\Gamma (s)} ~_2F_1 \left(-\frac 1 2,s-\frac 1
2;\frac 1 2;-\left(\frac{l+\frac 1 2}{mR}\right)^2\right)\right.\nn\\
& &\hspace{4cm}\left.-\frac{l+\frac 1 2} 2 \mzs \right],\nn
\eeq
it turns out that
we may interchange the $\sum_l$ and the integral in Eq.~(\ref{eq:3.13})
{\it only} if for the real part $\Re \cac$ of the contour the
condition $\Re \cac <-1$ is satisfied. However, the argument $\Gamma
(-1/2+t) \Gamma
(s-1/2)/\Gamma (1/2+t)$ has a pole at $t=1/2$. Thus the contour $\cac$
coming from $-i\infty$ must cross the real axis to the right of
$t=1/2$, and then
once more between $0$ and $1/2$ (in order that the pole $t=0$ of $\Gamma
(-t)$ lies to the right of it), before going to $+i\infty$. That is,
before
interchanging the sum and the integral we have to shift the contour
$\cac$ over the pole at
$t=1/2$ to the left, cancelling the (potentially divergent) second piece
in $A_{-1}^D (s)$. Closing then the contour to the left, we end up
with the following expression in terms of Hurwitz zeta functions
\beq
A_{-1}^D (s) =\frac{\rzs}{2\sqrt{\pi}\Gamma (s)}\sujnu \lll
(mR)^{2j}\frac{\g j+s-\frac 1 2\right)}{s+j}\zeta_H (2j+2s
-2;1/2).\label{eq:3.15}
\eeq
For $A_0^D$ one only needs to use the binomial expansion in order to find
\beq
A_0^D (s) =-\frac{\rzs}{2\Gamma (s)}\sujnu \lll (mR)^{2j}\Gamma (s+j)
\zeta_H (2j+2s-1;1/2).\label{eq:3.16}
\eeq
The series are convergent for $|mR|<1/2$. These representations
(\ref{eq:3.15}) and (\ref{eq:3.16}) show very clearly the analytic
structure
of $A_{-1}^D(s)$ and $A_0^D (s)$. As the {\it fifth} (and final) {\it
step} of our procedure, we are left with the quite simple task of explictly
evaluating this analytic structure, namely of finding its poles and
 some point
values, and of adding all contributions together.

The point values $A_{-1,0}^D(-p)$,
$p\in\nats_0$ ---respectively their residues in $s=1/2,-(2l+1)/2$,
$l\in\nats_0$--- necessary for the calculation of the associated
heat-kernel coefficients are immediate to obtain, using just
\beq
\zeta_H (1+\ep ,1/2) &=&\frac 1 {\ep} +\cao (\ep ^0),\nn\\
\Gamma (\ep -n) &=& \frac 1 {\ep} \frac{(-1)^n}{n!} +\cao (\ep
^0).\label{eq:3.16a}
\eeq
However, before we can actually calculate (an in principle arbitrary
number of) the heat-kernel coefficients, we need to obtain analytic
expressions for the
$A_i^D(s)$, $i\in\nats$. As is easy to see, they are
similar to the
ones for $A_{-1}^D (s)$ and $A_0^D(s)$ above. We need to
recall only that $D_i (t)$, Eq.~(\ref{eq:3.6}), is a polynomial in $t$,
\beq
D_i (t) =\suani x_{i,a} t^{i+2a},\label{eq:3.17}
\eeq
which coefficients $x_{i,a}$ are easily found by using
Eqs.~(\ref{eq:3.5}) and (\ref{eq:3.6}) directly, or either by using the
direct recursion
relations presented in appendix A. Thus
the calculation of $\aid$ is essentially solved through the identity
\beq
\iinma t^n &=& -\mzs \frac{n}{2(mR)^n}\frac{\g s+\frac n 2 \right)
\Gamma (1-s)}{\g 1+\frac n 2\right)}\nn\\
& &\qquad\qquad \times\nu ^n \left[1+\numr
\right]^{-s-\frac n 2}.\label{eq:3.18} \eeq
The remaining sum may be done as mentioned for $A_0^D$, and we end up
with \beq
A_i^D (s) &=& -\frac{\rzs}{\Gamma (s)}\sujnu \lll (mR)^{2j} \zeta_H
(-1+i+2j+2s;1/2) \nn\\
& &\hspace{3cm} \times \suani x_{i,a} \frac{(i+2a)\g s+a+j+\frac i
2\right)}{\g 1+a+\frac i 2\right)}.\label{eq:3.19}
\eeq

In summary we have obtained the analytic expression of all the
asymptotic terms
coming from  expansion (\ref{eq:3.4}) in its most elementary form,
which involves the very familiar Hurwitz zeta functions and Gamma
functions only. Expressions
(\ref{eq:3.15}), (\ref{eq:3.16}) and (\ref{eq:3.19}) constitute the
explicit starting
point for the calculation of an ---in principle arbitrary--- number of
heat-kernel coefficients in an extremely quick way.

\section{Heat-kernel coefficients for Dirichlet boundary conditions on
the three-dimensional ball}
\setcounter{equation}{0}
Let us now see how the analysis in Sect.~2 can be used for a very
effective calculation of the heat-kernel coefficients. The
dependence of the coefficients on the mass is already contained in the
coefficients of the massless case through
\beq
K_m (t) =K_{m=0}(t) e^{-m^2t}\nn
\eeq
and for this reason we shall restrict ourselves to $m=0$. For the sums in
(\ref{eq:3.15}), (\ref{eq:3.16}) and (\ref{eq:3.19}) this means that
only $j=0$ will contribute.

We shall distinguish between the coefficients $B_k$ with integer and
half-integer index $k$, because the situation is actually different
in both cases. In fact, corresponding to Eq.~(\ref{residuum}) (resp.
Eq.~(\ref{fusvalue})), the residue of (resp. the value of the function)
$\zeta_D$ is needed.

Let us start with the case of integer index $k\in\nats$, so that
Res$\,\,\zeta_D
(3/2 -k)$ is to be calculated. In order that $Z_D(s)$ does
not contribute,
one has to choose $N=2k-1$ and thus only the asymptotic terms $A_j^{
D}(s)$, $j=-1,0,1,...,2k-1$, will provide some contribution.
Furthermore, one may see
very easily which terms in the different $A_j^D(s)$ contribute. An
important
feature is, that for $i=2n$, $n\in\nats_0$, $A_i^D(s)$ does not
contribute
to $B_k$ for $k\in\nats$. The relevant residues are found to be
\beq
{\rm Res}\,\,A_{-1}^D\left(\frac 3 2 -k\right)&=&
\frac{(-1)^{k-1}}{(k-1)!}\frac{R^{3-2k}}{2\sqrt{\pi}\Gamma\left(\frac 5
2 -k\right)}\zeta_H \left(1-2k;\frac 1 2\right),\nn\\
{\rm Res}\,\,A_{2k-1}^D\left(\frac 3 2-k\right)&=& -\frac{R^{3-2k}}
{2\Gamma\left(\frac 3 2 -k\right)
}\sum_{a=0}^{2k-1}x_{2k-1,a}\frac{(2k-1+2a)a!}{\Gamma\left(\frac 1 2
+a+k\right)},\nn
\eeq
and for $n\in\nats$, $n\leq k-1$, $k\leq 3n$,
\beq
{\rm Res}\,\,A_{2n-1}\left(\frac 3 2 -k\right)
=\frac{(-1)^kR^{3-2k}}{\Gamma\left(\frac 3 2 -k\right)}\zeta_H \left(
1+2n-2k;\frac 1 2\right)
\sum_{a=0}^{k-1-n}x_{2n-1,a}\frac{(-1)^{a+n}(2n+2a-1)}{(k-1-a-n)!},\nn
\eeq
whereas for $n\leq k-1$, $k>3n$, we have
\beq
{\rm Res}\,\,A_{2n-1}\left(\frac 3 2 -k\right)
=\frac{(-1)^kR^{3-2k}}{\Gamma\left(\frac 3 2 -k\right)}\zeta_H \left(
1+2n-2k;\frac 1 2\right)
\sum_{a=0}^{2n-1}x_{2n-1,a}\frac{(-1)^{a+n}(2n+2a-1)}{(k-1-a-n)!}.\nn
\eeq
{}From these results we readily obtain the heat-kernel coefficients
through
\beq
{\rm Res}\,\,\zeta_D \left(\frac 3 2 -k\right) ={\rm Res}
\,\,\sum_{l=0}^k A_{2l-1}^D\left(\frac 3 2 -k\right)
\equiv \frac{B_k}{(4\pi)^{\frac 3 2} \Gamma\left(\frac 3 2
-k\right)}.\nn \eeq
The coefficients up to $B_{10}$ are listed in appendix B.

Let us now consider the calculation of the coefficients corresponding to
half-integer index $B_{k+1/2}$, $k\in\nats$. Here the value of
$\zeta_D (3/2-k)$ is needed and one finds $N=2k$.
It is apparent that the $A_i^D(s)$ with odd $i$, $i=2j-1$,
$j\in\nats_0$, do
not contribute now.
The relevant values of the $A_i^D(s)$
read
\beq
A_0^D(1-k)&=&-\frac{R^{2-2k}} 2 \zeta_H \left(1-2k;\frac 1
2\right),\nn\\
A_{2k}^D(1-k) &=& (-1)^k (k-1)! R^{2-2k}\sum_{a=0}^{2k}x_{2k,a}\frac{a!}
{(a+k-1)!},\nn
\eeq
and for $n\in\nats$, $n\leq k-1$, $k\leq 3n-1$,
\beq
A_{2n}^D(1-k) = -2R^{2-2k}(k-1)! \zeta_H \left( 1+2n-2k;\frac 1 2\right)
   \sum_{a=0}^{k-n-1}x_{2n,a} \frac{(-1)^{n+a}}{(k-n-a-1)!(a+n-1)!},\nn
\eeq
whereas for $n\leq k-1$, $k>3n-1$, we have
\beq
A_{2n}^D(1-k) = -2R^{2-2k}(k-1)! \zeta_H \left( 1+2n-2k;\frac 1 2\right)
   \sum_{a=0}^{2n}x_{2n,a} \frac{(-1)^{n+a}}{(k-n-a-1)!(a+n-1)!}.\nn
\eeq
And from these results, we finally obtain
\beq
\zeta_D (1-k) =\sum_{n=0}^k A_{2n}(1-k)
\equiv \frac{(-1)^{k-1}(k-1)!}{(4\pi)^{\frac 3 2}}B_{k+\frac 1 2}.\nn
\eeq
The heat-kernel coefficients $B_{k+\frac 1 2}$ are listed in
appendix B too.
Using $\zeta_H (-n;q)=-B_{n+1}(q)/(n+1)$,
$n\in\nats_0$, the results might have been given, equivalently, in terms of
Bernoulli polynomials $B_{n+1}(q)$.

\section{Robin boundary conditions on the three-dimensional ball}
\setcounter{equation}{0}
When Robin boundary conditions are imposed, using the same method of
the preceding sections  we can write the zeta function as
\beq
\zeta_R(s) =\sulnu (2l+1) \ikma\ln \left[\frac u R J_{l+\frac 1 2} (kR)
+kJ'_{l+\frac 1 2} (kR)\right] \label{eq:5.1} \eeq
and, in analogy with Eq.~(\ref{eq:3.3}), we then consider
\beq
\zeta_R^{\nu} = \sip \int\limits_m^{\infty}dk\,\, [k^2-m^2]^{-s}
\frac{\partial}{\partial k}
\ln \left[k^{-\nu
} (\frac u R I_{\nu} (kR) +kI'_{\nu} (kR))\right].\label{eq:5.2} \eeq
Employing the same idea as for Dirichlet boundary conditions,
this time we have in addition the following uniform asymptotic expansion
\cite{abramowitzstegun72} \beq
I'_{\nu} (\nu z) \sim \frac 1 {\sqrt{2\pi \nu}}\frac{e^{\nu
\eta}(1+z^2)^{\frac 1 4}} z \left[1+\sum_{k=1}^{\infty} \frac{v_k (t)}
{\nu ^k}\right],\label{eq:5.3}
\eeq
with the $v_k(t)$ determined by
\beq
v_k (t) =u_k (t) +t(t^2-1) \left[ \frac 1 2 u_{k-1}(t)
+tu'_{k-1}(t)\right]. \nn \eeq
In analogy with Eq.~(\ref{eq:3.6}), we write
\beq
\ln\left[1+\sum_{k=1}^{\infty}\frac{v_k (t)}{\nu
^k}+\frac{u}{\nu}t\left(1+\sum_{k=1}^{\infty}\frac{u_k (t)}{\nu
^k}\right)\right] \sim \sum_ {n=1}^{\infty}\frac{M_n (t)}{\nu
^n},\label{eq:6.1}
\eeq
where the functions $M_n (t)$ are easily obtained.
At this point we see
already, that for Robin boundary conditions no additional
calculation is necessary. Comparing the expansion (\ref{eq:5.3}) with
(\ref{eq:3.4}) and introducing $A_i ^R (s)$ for the contributions
coming from the asymptotic terms, one has
\beq
A_{-1}^R(s) =A_{-1}^D(s),\qquad A_0^R(s)=-A_0^D (s).\label{eq:5.5}
\eeq
Furthermore, the functions
$M_i (t)$ are of the form
\beq
M_i (t) =\suanzi z_{i,a} t^{i+a} \label{eq:6.2}
\eeq
(notice that here, in contrast with the case of Dirichlet boundary
 conditions, all
powers between $i$ and $3i$ are present). As a result, we find
\beq
A_i^M (s) &=& -\frac{\rzs}{\Gamma (s)}\sujnu \lll (mR)^{2j} \zeta_H
(-1+i+2j+2s;1/2) \nn\\
& &\hspace{3cm} \times \suanzi z_{i,a} \frac{(i+a)\g s+j+\frac {i+a}
2\right)}{\g 1+\frac {i+a} 2\right)}.\label{eq:6.3}
\eeq
One can show again that only the even indices $i$ contribute to the
residues of $\zeta_R (s)$, whereas the odd ones will contribute to the
point values.

Restricting ourselves as before (see the comment in the previous
section) to the massless case, the results for the heat-kernel
coefficients may now be read off from the formulas in the previous
section. One has
\beq
{\rm Res}\,\, A_{-1}^R \left(\frac 3 2 -k\right) &=& {\rm Res}\,\,
A_{-1}^D \left(\frac 3 2 -k\right),\nn\\
{\rm Res}\,\,A_{2k-1}^R \left(\frac 3 2 -k\right) &=&
-\frac{R^{3-2k}}{2\Gamma\left(\frac 3 2
-k\right)}\sum_{a=0}^{4k-2}z_{2k-1,a}\frac{(2k-1+a)\Gamma\left(1+\frac a
2\right)}{\Gamma\left(\frac 1 2 +k+\frac a 2\right)}, \nn
\eeq
where the expressions for $A_{2n-1}^R$ are found from the results in
Sect. 3,
once $x_{2n-1,a}$ has been replaced with $z_{2n-1,2a}$.
The coefficients for Neumann boundary conditions are given in appendix
C, and for the general case ($u$ arbitrary) in appendix D.
For the point values the analogous formulas read
\beq
A_0^R (1-k) &=& -A_0^D (1-k), \nn\\
A_{2k}^R (1-k) &=& (-1)^kR^{2-2k} (k-1)! \sum_{a=0}^{4k} z_{2k,a}
\frac{(2k+a)\Gamma\left(1+\frac a 2\right)}{2\Gamma\left(1+k+\frac a
2\right)}, \nn
\eeq
and once more the replacement of $x_{2n,a}$ with $z_{2n,2a}$ leads to the
results for $A_{2n}^R$. The results for the heat-kernel coefficients are
summarized in Apps. C and D.

\section{Generalization to the $D$-dimensional ball}
As we will now explain, for the generalization of our results to the
case of a $D$-dimensional ball almost no
additional calculations are necessary. Let us discuss first the case of
Dirichlet boundary condition. The starting point of the analysis is now
\beq
\zeta_D(s) =\sulnu d_l (D) \ikma \ln J_{l+\frac{D-2} 2} (kR).
\label{eq:7.1}
\eeq
It is easy to see that the above treatment for the individual terms of the
$l$-series,
 \beq
\zend = \ikma \ln J_{\nu} (kR)\label{eq:7.2}
\eeq
remains valid, once we have set $\nu =l+(D-2)/2$. In order to use our
procedure for the whole $l$-summation, what remains to be done is
to substitute for
the degeneracy $d_l (D)$ its value in powers of $l+(D-2)/2$, in order to
find again
expressions in terms of the Hurwitz zeta function
$\zeta_H (s;(D-2)/2)$. Writing
\beq
d_l (D) =\sual e_\alpha (D)
\left(l+\frac{D-2}2\right)^{\alpha},\label{eq:7.3} \eeq
the final results for $A_{-1}^D(s)$, $A_0^D(s)$ and $A_i^D(s)$,
$i\in\nats$, may be
read off from Eqs.~(\ref{eq:3.15}), (\ref{eq:3.16}) and (\ref{eq:3.19}).
We find
\beq
A_{-1}^D (s) &=&\frac{\rzs}{4\sqrt{\pi}\Gamma (s)}\sujnu \lll
(mR)^{2j}\frac{\g j+s-\frac 1 2\right)}{s+j}\label{eq:7.4}\\
& &\times \left[\sual \ead\zeta_H (2j+2s -1-\alpha;(D-2)/2)\right],\nn\\
A_0^D (s) &=&-\frac{\rzs}{4\Gamma (s)}\sujnu \lll (mR)^{2j}\Gamma
(s+j)\label{eq:7.5}\\
& &\times \left[\sual \ead\zeta_H (2j+2s-\alpha;(D-2)/2)\right],\nn\\
A_i^D (s) &=& -\frac{\rzs}{2\Gamma (s)}\sujnu \lll
(mR)^{2j}\label{eq:7.6}\\
& &\hspace{1cm} \times \left[\sual \ead\zeta_H
(-\alpha+i+2j+2s;(D-2)/2)\right] \nn\\
& &\hspace{2cm} \times \suani x_a \frac{(i+2a)\g s+a+j+\frac i
2\right)}{\g 1+a+\frac i 2\right)}.\nn
\eeq
We shall spare the reader the analogous results for Robin boundary
conditions. They
need not be given explicitly, since the procedure is absolutely clear  by
now. Let us just write down the relevant residues and point values of
$\zeta_D
(s)$ (the Robin case follows from the replacements explained in Sect.
5). They read
\beq
{\rm Res}\,\,A_{-1}^D\left(\frac 3 2 -k\right) &=&
\frac{(-1)^{k-1}}{(k-1)!} \frac{R^{3-2k}}{4\sqrt{\pi}\Gamma\left(\frac 5
2 -k\right)}\sual \ead \, \zeta_H \left(2-2k-\alpha;\frac{D-2}
2\right), \nn
\eeq
for $n=1,...,k-1$, $k>3n$,
\beq
{\rm Res}\,\,A_{2n-1} \left(\frac 3 2 -k\right) &=&(-1)^k
\frac{R^{3-2k}}{2\Gamma\left(\frac 3 2 -k\right)}\sual \ead \, \zeta_H
\left( 2+2n-\alpha -2k; \frac{D-2}2\right)\nn\\
& &\qquad \times
\sum_{a=0}^{2n-1}(-1)^{n+a}x_{2n-1,a}\frac{(2n+2a-1)}{(k-1-n-a)!
\Gamma\left(\frac 1 2 +a+n\right)},\nn
\eeq
whereas for $k\leq n$, it reads
\beq
{\rm Res}\,\,A_{2n-1} \left(\frac 3 2 -k\right) &=&(-1)^k
\frac{R^{3-2k}}{2\Gamma\left(\frac 3 2 -k\right)}\sual \ead \, \zeta_H
\left( 2+2n-\alpha -2k; \frac{D-2}2\right)\nn\\
& &  \qquad \times
\sum_{a=0}^{k-n-1}(-1)^{n+a}x_{2n-1,a}\frac{(2n+2a-1)}{(k-1-n-a)!
\Gamma\left(\frac 1 2 +a+n\right)}.\nn
\eeq
For higher indices it is adviceable to distinguish between $D$ even and
$D$
odd. For $D$ odd contributions arise for $n=k,...,k+(D-3)/2$, and read
\beq
{\rm Res}\,\,A_{2n-1} \left(\frac 3 2 -k\right) &=&-
\frac{R^{3-2k}}{4\Gamma\left(\frac 3 2 -k\right)}e_{1+2n-2k}
\sum_{a=0}^{2n-1}x_{2n-1,a}\frac{(2n+2a-1)(a+n-k)!}
{\Gamma\left(\frac 1 2 +a+n\right)},\nn
\eeq
whereas for $D$ even the indices run from $n=k,...,k+(D-4)/2$, and the
results are
\beq
{\rm Res}\,\,A_{2n} \left(\frac 3 2 -k\right) &=&-
\frac{R^{3-2k}}{2\Gamma\left(\frac 3 2 -k\right)}e_{2+2n-2k}
\sum_{a=0}^{2n}x_{2n,a}\frac{\Gamma\left(\frac 3 2 -k+a+n\right)}
{(a+n-1)!}.\nn
\eeq
Let us conclude with the list of point values. The leading
asymptotics $A_{-1}^D$ gives only contributions for $k=0$,
\beq
A_{-1} (0) =-\frac 1 2 \sual \ead \zeta_H \left(-\alpha -1;
\frac{D-2}2\right).\nn
\eeq
Furthermore, for $n=1,...,k-1$, we have
\beq
A_{2n} \left(1 -k\right) &=&-
R^{2-2k}(k-1)!\sual \ead \zeta_H\left(-\alpha
+2n+2-2k;\frac{D-2}2\right)\nn\\
& &\qquad \times \sum_{a=0}^{2n}x_{2n,a}\frac{(-1)^{a+n}}
{(a+n-1)!(k-1-a-n)!},\nn
\eeq
if $k>3n-1$, and if $k\leq 3n-1$
\beq
A_{2n} \left(1 -k\right) &=&-
R^{2-2k}(k-1)!\sual \ead \, \zeta_H\left(-\alpha
+2n+2-2k;\frac{D-2}2\right)\nn\\
& &\qquad \times \sum_{a=0}^{k-n-1}x_{2n,a}\frac{(-1)^{a+n}}
{(a+n-1)!(k-1-a-n)!}.\nn
\eeq
Finally, for $D$ odd and for $n=k,...,k+(D-3)/2$,
\beq
A_{2n} \left(1 -k\right) &=&\frac 1 2 (-1)^k(k-1)!
R^{2-2k}e_{1+2n-2k}
\sum_{a=0}^{2n}x_{2n,a}\frac{(a+n-k)!}
{(a+n-1)!},\nn
\eeq
whereas for $D$ even the indices run from $n=k+1,...,k+(D-2)/2$, and the
result reads
\beq
A_{2n-1} \left(1 -k\right) &=&\frac 1 4 (-1)^k(k-1)!
R^{2-2k}e_{2n-2k}
\sum_{a=0}^{2n-1}x_{2n-1,a}\frac{(2n-1+2a)\Gamma\left(\frac 1 2
-k+a+n\right)} {\Gamma\left(\frac 1 2+ a+n\right)}.\nn
\eeq
The formulas above simplify a bit if we write
the degeneracy (\ref{eq:2.4}) under the form
\beq
d_l (D) &=& \frac 2 {(D-2)!}\left[\left(l+\frac{D-2}
2\right)^2-\left(\frac D 2
-2\right)^2\right]\times...\times\left(l+\frac{D-2}2\right),\;\;{\rm
for}\phantom{a}D\phantom{a}{\rm odd},\nn\\
d_l (D) &=& \frac 2 {(D-2)!}\left[\left(l+\frac{D-2}
2\right)^2-\left(\frac D 2
-2\right)^2\right]\times...\times\left(l+\frac{D-2}2\right)^2,\;\;{\rm
for}\phantom{a} D\phantom{a}{\rm even},\nn
\eeq
so that $e_{2k}(D) =0$ for $D$ odd and $e_{2k-1}(D) =0$ for $D$ even,
$k\in\nats$.

Furthermore, one might use the following recursion for the coefficients
$\ead$ appearing in the expression of the degeneracy $d_l(D)$,
Eq.~(\ref{eq:7.3}),
\beq
e_{2\alpha} (D+2) &=& \frac 1 {D(D-1)}\left[e_{2\alpha -2} (D)
-\left(\frac D 2 -1\right)^2 e_{2\alpha} (D)\right], \;\;{\rm
for}\phantom{a}D\phantom{a}{\rm even},\nn\\
e_{2\alpha-1} (D+2) &=& \frac 1 {D(D-1)}\left[e_{2\alpha -3} (D)
-\left(\frac D 2 -1\right)^2 e_{2\alpha-1} (D)\right],  \;\;{\rm
for}\phantom{a} D\phantom{a} {\rm odd},\nn
\eeq
where we have used the definitions $e_{-k}(D) =0$ for $k\in\nats_0$ and
$e_{\alpha} (D) =0$ for $\alpha >D-2$.

We have performed explicit calculations for $D=4$ and $D=5$. One
has in these cases
\beq
d_l (4) &=& (l+1)^2, \qquad e_1(4) =0,\,\, e_2(4) =1,\nn\\
d_l (5) &=& \frac 1 3 \left(l+\frac 3 2 \right) \left[\left( l+\frac 3 2
\right)^2 -\frac 1 4 \right],\qquad e_1 (5) =-\frac 1 {12},\,\,e_2 (5)
=0,\,\, e_3 (5) =\frac 1 3.\nn
\eeq
The results for the heat-kernel coefficients are presented in Apps.
B, C and D.

\section{Conclusions}

As promised in the introduction, we have developed in this paper a very
convenient method in order to deal with the problem of the calculation of
 heat-kernel
coefficients corresponding to an arbitrary elliptic operator with any of
the usual boundary conditions (Dirichlet, Neumann or Robin),
with the only proviso that the behavior of some basis for its spectrum
should be known (even if the eigenvalues themselves are actually unknown).

This is indeed a very common case in mathematical physics, what
 conferes to our procedure a
wide generality of application.
Another fundamental characteristic of the method is
its extreme simplicity, which comes in part from the quite strong
background on
zeta function computations that we have acquired during the last half
a dozen years. This knowledge conferes to the new method the same
elegance that the procedure of zeta function regularization (including
the analytic continuation techniques and non-trivial series commutation
that it involves) has in itself.

Finally, we have tried our method with explicit
examples and give several tables of heat-kernel coefficients that
have been calculated here (with relative  easiness) for the first time.
For the near future we envisage to investigate other physical applications
where the method can prove useful.

\vspace{5mm}

{\it Note:} At the final stage of our analysis, P.~Gilkey made us aware
of related research by M.~Levitin \cite{ml}, who has further developed
the approach of Kennedy \cite{kennedy78,kennedy78a}, also with
the aim of calculating higher-order heat-kernel coefficients. We are
indebted with M.~Levitin for sending us his results, which have served as
a very good
check of our calculations. All results in common with his are in complete
agreement.

\vspace{5mm}

\noindent{\large \bf Acknowledgments}

It is a pleasure to thank S.~Dowker, G.~Esposito, P.~Gilkey,
S.~Leseduarte and especially M.~Levitin for interesting discussions and
helpful comments.
K.K. thanks the members of the Department ECM of the University of
Barcelona for their warm hospitality, and acknowledges financial support
from the Alexander von Humboldt Foundation (Germany).
This work has been supported by DGICYT (Spain), project Nos.
PB93-0035 and HA93-004, by CIRIT (Generalitat de Catalunya), and by DAAD
(Acciones Integradas).

\newpage

\begin{appendix}
\section{Recursion relation for the coefficients $x_{i,a}$}
\setcounter{equation}{0}
In this appendix we present the recursion relations for the coefficients
$x_{i,a}$, Eq.~(\ref{eq:3.17}). For convenience let us introduce for
$i\in\nats$, $a=0,...,i$,
\beq
x_{i,a}=\frac{c_{i+1,a}}{2^{i+1}(i+2a)}.\nn
\eeq
Then, starting with $c_{1,0}=-1$, we find the following recursion
relation,
\[c_{i,0}=(i-2)c_{i-1,0}-\frac{1}{2}\sum_{s=1}^{i-1}c_{i-s,0}c_{s,0},\]
\[c_{i,i-1}=(4-3i)c_{i-1,i-2}+\frac{1}{2}\sum_{s=1}^{i-1}c_{i-s,i-s-1}
c_{s,s-1},\]
and for $a=1, ... ,i-2$, we have
\begin{eqnarray}c_{i,a}&=&(i-2+2a)(c_{i-1,a}-c_{i-1,a-1})\nonumber\\
&&-\frac{1}{2}\sum_{s=1}^{i-1}
\left(\sum_{j=Max(0,1+a+s-i)}^{Min(a,s-1)}c_{i-s,a-j}c_{s,j}-
\sum_{j=Max(0,a+s-i)}^{Min(a-1,s-1)}c_{i-s,a-j-1}c_{s,j}\right).
\nonumber\end{eqnarray}
This relation can be used very effectively for the calculation of the
coefficients $x_{i,a}$.
\renewcommand{\theequation}{{\mbox B}.\arabic{equation}}
\section{Heat-kernel coefficients for Dirichlet boundary conditions}
\setcounter{equation}{0}
In this appendix we list our results for the heat-kernel coefficients of
the Laplace operator in $3,4$ and $5$ dimensions with Dirichlet boundary
conditions. Here and in the following appendices, the first
coefficients $B_0,...,B_{5/2}$ are listed for completeness and may also
be found in \cite{kennedy78,kennedy78a} or derived from
\cite{bransongilkey90}.

In three dimensions we have found that
\beq
B_0&=& \frac 4 3 \pi R^3 \nn\\
B_{1/2}&=&-2\pi^{3/2}R^2\nn\\
B_1&=&\frac{8\pi R}3 \nn\\
B_{3/2}&=&-\frac 1 6 \pi^{3/2}\nn\\
B_2&=& -\frac{16\pi}{315R} \nn\\
B_{5/2}&=&-\frac{\pi^{3/2}}{120R^2}\nn\\
B_3 &=& -\frac{64\pi}{9009R^3} \nn\\
B_{7/2} &=& -\frac{47\pi^{3/2}}{20160R^4}\nn\\
B_4 &=& -\frac{202816\pi}{72747675R^5} \nn\\
B_{9/2} &=& -\frac{521\pi^{3/2}}{443520R^6}\nn\\
B_5 &=& -\frac{25426048\pi}{15058768725R^7} \nn\\
B_{11/2} &=& -\frac{9521\pi^{3/2}}{11531520R^8}\nn\\
B_6 &=& -\frac{90878576896\pi}{67689165418875R^9}  \nn\\
B_{13/2} &=& -\frac{34344493\pi^{3/2}}{47048601600R^{10}}\nn\\
B_7 &=& -\frac{22835854180352\pi}{17531493843488625R^{11}}  \nn\\
B_{15/2} &=& -\frac{36201091\pi^{3/2}}{47048601600R^{12}}\nn\\
B_8 &=&
-\frac{1509389910845640704\pi}{1019964780320324713875R^{13}} \nn\\
B_{17/2} &=&
-\frac{153984929039\pi^{3/2}}{164481911193600R^{14}}\nn\\
B_{9}
&=&
-\frac{1673450232605639069696\pi}{872477873086005760248675R^{15}} \nn\\
B_{19/2} &=&
-\frac{13334525091737\pi^{3/2}}{10362360405196800R^{16}}\nn\\
B_{10} &=&
-\frac{643985013732181345325056\pi}{231206636367791526465898875R^{17}}.\nn
\eeq
In four dimensions the result is
\beq
B_0&=&\frac 1 2 \pi^2R^4\nn\\
B_{1/2}&=&-\pi^{5/2}R^3\nn\\
B_1&=&2\pi^2R^2\nn\\
B_{3/2}&=&-\frac{11\pi^{5/2}R}{32}\nn\\
B_2&=&-\frac{4\pi^2}{45}\nn\\
B_{5/2}&=&-\frac{35\pi^{5/2}}{4096R}\nn\\
B_{3} &=&- \frac{464\pi^2}{45045R^2}\nn\\
B_{7/2}& =& -\frac{911\pi^{5/2}}{196608R^3}\nn\\
B_{4}& =& -\frac{107456\pi^2}{14549535R^4}\nn\\
B_{9/2}& =& -\frac{827315\pi^{5/2}}{201326592R^5}\nn\\
B_{5}& =& -\frac{23288576\pi^2}{3011753745R^6}\nn\\
B_{11/2}& =& -\frac{158590273\pi^{5/2}}{32212254720R^7}\nn\\
B_{6} &=& -\frac{20064545792\pi^2}{1933976154825R^8}\nn\\
B_{13/2}& =& -\frac{630648945109\pi^{5/2}}{86586540687360R^9}\nn\\
B_{7} &=& -\frac{492912963584\pi^2}{29464695535275R^{10}}\nn\\
B_{15/2}& =&
-\frac{70309732006867\pi^{5/2}}{5541538603991040R^{11}}\nn\\
B_{8} &=& -\frac{37648078688043008\pi^2}{1204208713483264125R^{12}}\nn\\
B_{17/2}& =&
-\frac{1578924180477650401\pi^{5/2}}{62419890835355074560R^{13}}\nn\\
B_{9} &=&
-\frac{887504373820227584\pi^2}{13409327181833639595R^{14}}\nn\\
B_{19/2}& =& -\frac{1018264365864160946171\pi^{5/2}}
                     {17976928560582261473280R^{15}}\nn\\
B_{10}& =&
-\frac{252629551155828479492096\pi^2}{1616829624949591094167125R^{16}}.
\nn \eeq
Finally, in five dimensions we obtain
\beq
B_0&=&\frac{8\pi^2R^5}{15}\nn\\
B_{1/2}&=&-\frac{4\pi^{5/2}R^4}3\nn\\
B_1&=&\frac{32\pi^2R^3}9\nn\\
B_{3/2}&=&-\pi^{5/2}R^2\nn\\
B_2&=&-\frac{128\pi^2R}{945}\nn\\
B_{5/2}&=&\frac{17\pi^{5/2}}{360}\nn\\
B_{3}& =& \frac{1216\pi^2}{45045R}\nn\\
B_{7/2}& =& \frac{157\pi^{5/2}}{30240R^2}\nn\\
B_{4}& =& \frac{235264\pi^2}{43648605R^3}\nn\\
B_{9/2}& =&\frac{5\pi^{5/2}}{2464R^4}\nn\\
B_{5}& =& \frac{779264\pi^2}{280598175R^5}\nn\\
B_{11/2} &=& \frac{593\pi^{5/2}}{449280R^6}\nn\\
B_{6}& =& \frac{91757946368\pi^2}{43074923448375R^7}\nn\\
B_{13/2} &=& \frac{32815499\pi^{5/2}}{28229160960R^8}\nn\\
B_{7}& =& \frac{22103738934272\pi^2}{10518896306093175R^9}\nn\\
B_{15/2}& =& \frac{119034319\pi^{5/2}}{94097203200R^{10}}\nn\\
B_{8}& =& \frac{53300366610079744\pi^2}{21397862524202616375R^{11}}\nn\\
B_{17/2}& =& \frac{798608979601\pi^{5/2}}{493445733580800R^{12}}\nn\\
B_{9}& =&
\frac{381809787573414866944\pi^2}{111856137575128943621625R^{13}}\nn\\
B_{19/2}& =&
\frac{146801666871373\pi^{5/2}}{62174162431180800R^{14}}\nn\\
B_{10} &=&
\frac{31815282789579439112192\pi^2}{6031477470464126777371275R^{15}}\nn
\eeq
\section{Heat-kernel coefficients for Neumann boundary conditions}
\setcounter{equation}{0}
Here is a list of the results we have obtained for the heat-kernel
coefficients of
the Laplace operator in $3,4$ and $5$ dimensions with Neumann boundary
conditions. In three dimensions we have found
\beq
B_0&=& \frac 4 3 \pi R^3\nn\\
B_{1/2}&=&2\pi^{3/2}R^2\nn\\
B_1&=&\frac{8\pi R}3\nn\\
B_{3/2}&=&\frac 7 6 \pi^{3/2}\nn\\
B_2&=& \frac{16\pi}{9R}\nn\\
B_{5/2}&=&\frac{47\pi^{3/2}}{60R^2}\nn\\
B_3& =& \frac{6464\pi}{6435R^3}\nn\\
B_{7/2}& =& \frac{3973\pi^{3/2}}{10080R^4}\nn\\
B_4& =& \frac{14766656\pi}{31177575R^5}\nn\\
B_{9/2} &=& \frac{5057\pi^{3/2}}{28160R^6}\nn\\
B_5& =& \frac{2314167424\pi}{10756263375R^7}\nn\\
B_{11/2}& =& \frac{2320069\pi^{3/2}}{27675648R^8}\nn\\
B_6& =& \frac{1439468204288\pi}{13537833083775R^9}\nn\\
B_{13/2}& =& \frac{11298472831\pi^{3/2}}{250925875200R^{10}}\nn\\
B_7& =& \frac{369968178163712\pi}{5843831281162875R^{11}}\nn\\
B_{15/2}& =& \frac{1718717967893\pi^{3/2}}{57211099545600R^{12}}\nn\\
B_8& =&
\frac{48366532825354366976\pi}{1019964780320324713875R^{13}}\nn\\
B_{17/2}& =&
\frac{113384991528329\pi^{3/2}}{4511503849881600R^{14}}\nn\\
B_9& =
&\frac{781980237125923045376\pi}{17805670879306240005075R^{15}}\nn\\
B_{19/2}& =&
\frac{33839928581307889\pi^{3/2}}{1326382131865190400R^{16}}\nn\\
B_{10}& =&
\frac{14392436216775440050663424\pi}{297265675330017676884727125
R^{17}}\nn
\eeq
In four dimensions
\beq
B_0&=&\frac 1 2 \pi^2R^4\nn\\
B_{1/2}&=&\pi^{5/2}R^3\nn\\
B_1&=&2\pi^2R^2\nn\\
B_{3/2}&=&\frac{41\pi^{5/2}R}{32}\nn\\
B_2&=&\frac{116\pi^2}{45}\nn\\
B_{5/2}&=&\frac{5861\pi^{5/2}}{4096R}\nn\\
B_3 &=& \frac{99472\pi^2}{45045R^2}\nn\\
B_{7/2}&=& \frac{388657\pi^{5/2}}{393216R^3}\nn\\
B_4 &=& \frac{18334144\pi^2}{14549535R^4}\nn\\
B_{9/2} &=& \frac{91095533\pi^{5/2}}{201326592R^5}\nn\\
B_5 &=& \frac{6269294336\pi^2}{15058768725R^6}\nn\\
B_{11/2} &=& \frac{2096614963\pi^{5/2}}{32212254720R^7}\nn\\
B_6 &=& -\frac{1448614636544\pi^2}{13537833083775R^8}\nn\\
B_{13/2} &=& -\frac{13041149176631\pi^{5/2}}{86586540687360R^9}\nn\\
B_7 &=& -\frac{38509398708224\pi^2}{100179964819935R^{10}}\nn\\
B_{15/2} &=& -\frac{1498787760061463\pi^{5/2}}
{5541538603991040R^{11}}\nn\\
B_8 &=&
-\frac{7562397933317668864\pi^2}{13246295848315905375R^{12}}\nn\\
B_{17/2} &=&
-\frac{23865356170241004641\pi^{5/2}}{62419890835355074560R^{13}}\nn\\
B_9 &=& -\frac{30045051913611575296\pi^2}
{36622112051226326625R^{14}}\nn\\
B_{19/2} &=&
-\frac{135252966433194092697787\pi^{5/2}}
{233700071287569399152640R^{15}}\nn\\
B_{10} &=&
-\frac{307843753219621367054336\pi^2}{230975660707084442023875R^{16}}
\nn
\eeq
And, finally, in five dimensions
\beq
B_0&=&\frac{8\pi^2R^5}{15}\nn\\
B_{1/2}&=&\frac{4\pi^{5/2}R^4}3\nn\\
B_1&=&\frac{32\pi^2R^3}9\nn\\
B_{3/2}&=&3\pi^{5/2}R^2\nn\\
B_2&=&\frac{1024\pi^2R}{135}\nn\\
B_{5/2}&=&\frac{1873\pi^{5/2}}{360}\nn\\
B_3 &=& \frac{63296\pi^2}{6435R}\nn\\
B_{7/2} &=& \frac{10121\pi^{5/2}}{1890R^2}\nn\\
B_4 &=& \frac{504064\pi^2}{61047R^3}\nn\\
B_{9/2} &=& \frac{198463\pi^{5/2}}{55440R^4}\nn\\
B_5 &=& \frac{125689856\pi^2}{30879225R^5}\nn\\
B_{11/2} &=& \frac{34154807\pi^{5/2}}{34594560R^6}\nn\\
B_6 &=& -\frac{56447170574848\pi^2}{157941385977375R^7}\nn\\
B_{13/2} &=& -\frac{16602940093\pi^{5/2}}{14114580480R^8}\nn\\
B_7& =& -\frac{945576485184512\pi^2}{281253911927625R^9}\nn\\
B_{15/2} &=& -\frac{13550828636809\pi^{5/2}}
{5721109954560R^{10}}\nn\\
B_8 &=& -\frac{259104011527854628864\pi^2}
{55634442562926802575R^{11}}\nn\\
B_{17/2} &=&
-\frac{5379580705269259\pi^{5/2}}{1973782934323200R^{12}}\nn\\
B_9 &=&
-\frac{46180677500935662030848\pi^2}{9587668935011052310425R^{13}}\nn\\
B_{19/2} &=&
-\frac{2640354677256557617\pi^{5/2}}{994786598898892800R^{14}}\nn\\
B_{10} &=&
-\frac{1401638457879249954799616\pi^2}
{306775722734796364174125R^{15}}\nn
\eeq
\section{Heat-kernel coefficients for Robin boundary conditions}
\setcounter{equation}{0}
We conclude our list of results with the leading coefficients for
general Robin boundary conditions for $D=3,4$ and $5$.

In three dimensions, we have found
\beq
B_0&=&\frac{4\pi R^3}3\nn\\
B_{1/2}&=&2\pi^{3/2}R^2\nn\\
B_1&=&-\frac{4\pi R} 3 (1+6u)\nn\\
B_{3/2}&=&\frac{\pi^{3/2}}6 (1+24u^2)\nn\\
B_2 &=& \frac{2\pi}{45R}(1-18u+60u^2-120u^3)\nn\\
B_{5/2}&=&\frac{\pi^{3/2}}{60R^2} (2-15u+60u^2-120u^3+120u^4)\nn\\
B_3 &=& \frac{\pi}{45045R^3}
(1633 - 12870u + 46904u^2 \nn\\
& &\qquad - 107536u^3 + 144144u^4 - 96096u^5)\nn\\
B_{7/2} &=& \frac{\pi^{3/2}}{10080R^4}
(151 - 1008u + 3612u^2\nn\\
& &\qquad  -  8400u^3 + 13440u^4 - 13440u^5
+ 6720u^6)\nn\\
B_4 &=& \frac{\pi}{436486050R^5}
(8243319 - 51363270u + 169826940u^2  \nn\\
& &\qquad      - 395830040u^3 + 676878800u^4 -
       835097120u^5\nn\\
& &\qquad+ 665121600u^6 - 266048640u^7)\nn\\
B_{9/2} &=& \frac{\pi^{3/2}}{1774080R^6}
(14639 - 80784u + 249304u^2\nn\\
& &\qquad     -  556600u^3 + 976800u^4 -
1330560u^5\nn\\
& &\qquad   + 1340416u^6 - 887040u^7 +
295680u^8)\nn\\
B_5 &=& \frac{\pi}{301175374500R^7}
(3517532467 - 17760354570u\nn\\
& &\qquad+ 49945523040u^2 -
       105573378240u^3 + 182023225440u^4\nn\\
& &\qquad  -  259648898880u^5 + 295543449600u^6 -
       252181862400u^7\nn\\
& &\qquad+ 142779436800u^8 - 40794124800u^9)\nn
\eeq
In four dimensions, the results read
\beq
B_0&=&\frac{\pi^2 R^4}2\nn\\
B_{1/2}&=&\pi^{5/2}R^3\nn\\
B_1&=&-2\pi^2 R^2  (1+2u)\nn\\
B_{3/2}&=&\frac{\pi^{5/2}R}{32} (9+32u+64u^2)\nn\\
B_2 &=& -\frac{4\pi^2}{45}(1+30u^3)\nn\\
B_{5/2}&=&-\frac{\pi^{5/2}}{4096R} (59-224u+2048u^3-4096u^4)\nn\\
B_3 &=&-\frac{16\pi^2}{45045R^2}
(75 - 286u + 286u^2 + 858u^3\nn\\
& &\qquad      - 3003u^4 + 3003u^5)\nn\\
B_{7/2} &=& -\frac{\pi^{5/2}}{393216R^3}
(5807 - 21024u + 29952u^2\nn\\
& &\qquad  +  7168u^3 - 110592u^4 +
196608u^5 -
        131072u^6)\nn\\
B_4 &=& -\frac{32\pi^2}{14549535R^4}
(11726 - 39368u + 62016u^2\nn\\
& &\qquad - 36176u^3 - 75582u^4 +
230945u^5\nn\\
& &\qquad      - 277134u^6 + 138567u^7)\nn\\
B_{9/2} &=& -\frac{\pi^{5/2}}{201326592R^5}
(2961171 - 9105152u + 14440448u^2\nn\\
& &\qquad - 13142016u^3 - 458752u^4 +
25427968u^5\nn\\
& &\qquad       - 46137344u^6 + 41943040u^7 -
16777216u^8)\nn\\ B_5 &=& -\frac{64\pi^2}{15058768725R^6}
(6419236 - 17976600u + 27448200u^2\nn\\
& &\qquad - 28336920u^3 + 14866740u^4 +
14709420u^5\nn\\
& &\qquad   - 49365705u^6 + 65189475u^7 -
47805615u^8\nn\\
& &\qquad      + 15935205u^9)\nn
\eeq
Finally, in five dimensions we have found
\beq
B_0&=&\frac{8\pi^2 R^5}{15}\nn\\
B_{1/2}&=&\frac{4\pi^{5/2}R^4}3\nn\\
B_1&=&-\frac{8\pi^2 R^3}9  (5+6u)\nn\\
B_{3/2}&=&\frac{\pi^{5/2}R^2}{3} (3+8u+8u^2)\nn\\
B_2 &=& \frac{4\pi^2R}{135}(-5+6u-60u^2-120u^3)\nn\\
B_{5/2}&=&\frac{\pi^{5/2}}{360} (-17-240u^2+480u^4)\nn\\
B_3 &=& \frac{2\pi^2}{135135R}
(87 + 13442u - 35464u^2 \nn\\
& &\qquad +61776u^3 + 48048u^4 - 96096u^5)\nn\\
B_{7/2} &=& \frac{\pi^{5/2}}{7560R^2}
(-88 + 483u - 1806u^2 +
       2940u^3\nn\\
& &\qquad - 1680u^4 - 3360u^5 + 3360u^6)\nn\\
B_4 &=& \frac{\pi^2}{43648605R^3}
(-539501 + 4050078u - 12086660u^2\nn\\
& &\qquad +  23878744u^3 - 23715952u^4 +
1478048u^5\nn\\
& &\qquad  +  26604864u^6 - 17736576u^7)\nn\\
B_{9/2} &=& \frac{\pi^{5/2}}{2661120R^4}
(-18927 + 99616u - 302720u^2\nn\\
& &\qquad     +  576048u^3 - 748704u^4 +
473088u^5\nn\\
& &\qquad  + 177408u^6 - 591360u^7 +
295680u^8)\nn\\
B_5 &=& \frac{\pi^2}{90352612350R^5}
(-935536567 + 4964319990u\nn\\
& &\qquad-13111462800u^2 +
       25019918880u^3\nn\\
& &\qquad- 34365190560u^4 +
       32451298368u^5\nn\\
& &\qquad    -12409401600u^6 -
       12609093120u^7\nn\\
& &\qquad +20397062400u^8 - 8158824960u^9).\nn
\eeq
This concludes our lists of explicit tables for the heat-kernel coefficients.
In the same way, results for any desired dimension $D$ are very easy to
obtain from the formulas in the text.
 \end{appendix}

\end{document}